\documentstyle[12pt]{article}
\newcommand{\decentmargins}{
	\oddsidemargin=0in
	\evensidemargin=0in
	\textwidth=6.5in              %  paper is 8.5in wide

	\headheight=0pt
	\headsep=0pt
	\topmargin=0in
	\textheight=9.0in              %  paper is 11.0in high
	}
\def\slashchar#1{\setbox0=\hbox{$#1$}           % set a box for #1
   \dimen0=\wd0                                 % and get its size
   \setbox1=\hbox{/} \dimen1=\wd1               % get size of /
   \ifdim\dimen0>\dimen1                        % #1 is bigger
      \rlap{\hbox to \dimen0{\hfil/\hfil}}      % so center / in box
      #1                                        % and print #1
   \else                                        % / is bigger
      \rlap{\hbox to \dimen1{\hfil$#1$\hfil}}   % so center #1
      /                                         % and print /
   \fi}                                         %
\newcommand{\bra}[1]{\mbox{$\left\langle{#1}\right|$}}
\newcommand{\ket}[1]{\mbox{$\left|{#1}\right\rangle$}}

\newcommand{\md}[1]{\mbox{d$#1$}} %math differential
\newcommand{\tr}{\mbox{Tr}}
\newcommand{\bold}[1]{\mbox{\boldmath$#1$}}
\newcommand{\EFI}{Enrico Fermi Institute, 5640 S. Ellis Ave., Chicago, Illinois
60637}

\newcommand{\eqn}[1]{Eq.~(\protect\ref{#1})}
\newcommand{\twoeqn}[2]{Eqs.~(\protect\ref{#1}) and (\protect\ref{#2})}

\newcommand{\fig}[1]{Fig.~\protect\ref{#1}}
\newcommand{\tabref}[1]{Table~\protect\ref{#1}}

\newcommand{\ntabref}[2]{Tables~\protect\ref{#1}--\protect\ref{#2}}
\newcommand{\secref}[1]{Section~\protect\ref{#1}}
\newcommand{\refref}[1]{Ref.~\protect\cite{#1}}
\newcommand{\refrefs}[1]{Refs.~\protect\cite{#1}}

\newcommand{\etal}{{\em et al.\ }}
\catcode`@=11
\def\citen#1{%
\edef\@tempa{\@ignspaftercomma,#1, \@end, }% ignore spaces in parameter list
\edef\@tempa{\expandafter\@ignendcommas\@tempa\@end}%
\if@filesw \immediate \write \@auxout {\string \citation {\@tempa}}\fi
\@tempcntb\m@ne \let\@h@ld\relax \let\@citea\@empty
\@for \@citeb:=\@tempa\do {\@cmpresscites}%
\@h@ld}
\def\@ignspaftercomma#1, {\ifx\@end#1\@empty\else
   #1,\expandafter\@ignspaftercomma\fi}
\def\@ignendcommas,#1,\@end{#1}
\def\@cmpresscites{%
 \expandafter\let \expandafter\@B@citeB \csname b@\@citeb \endcsname
 \ifx\@B@citeB\relax % undefined
    \@h@ld\@citea\@tempcntb\m@ne{\bf ?}%
    \@warning {Citation `\@citeb ' on page \thepage \space undefined}%
 \else%  defined
    \@tempcnta\@tempcntb \advance\@tempcnta\@ne
    \setbox\z@\hbox\bgroup % check if citation is a number:
    \ifnum\z@<0\@B@citeB \relax
       \egroup \@tempcntb\@B@citeB \relax
       \else \egroup \@tempcntb\m@ne \fi
    \ifnum\@tempcnta=\@tempcntb % Number follows previous--hold on to it
       \ifx\@h@ld\relax % first pair of successives
          \edef \@h@ld{\@citea\@B@citeB}%
       \else % compressible list of successives
          \edef\@h@ld{\hbox{--}\penalty\@highpenalty \@B@citeB}%
       \fi
    \else   %  non-successor--dump what's held and do this one
       \@h@ld \@citea \@B@citeB \let\@h@ld\relax
 \fi\fi%
 \let\@citea\@citepunct
}
\def\@citepunct{,\penalty\@highpenalty\hskip.13em plus.1em minus.1em}%
\def\@citex[#1]#2{\@cite{\citen{#2}}{#1}}%
\def\@cite#1#2{\leavevmode\unskip
  \ifnum\lastpenalty=\z@ \penalty\@highpenalty \fi % highpenalty before
  \ [{\multiply\@highpenalty 3 #1% % triple-highpenalties within list
      \if@tempswa,\penalty\@highpenalty\ #2\fi % and before note.
    }]\spacefactor\@m}
\decentmargins

\newcommand{\cn}{Collaboration}
\newcommand{\mh}{m_h}
\newcommand{\mhstar}{m_{h^*}}
\newcommand{\mhh}{m_{h^{(*)}}}
\newcommand{\zz}{z^{(*)}}
\newcommand{\wmax}{\w_{\mbox{\scriptsize max}}}
\newcommand{\w}{w} %This is dumb but it simplifies my life
\newcommand{\mH}{m_H}
\newcommand{\iw}{\xi(\w)}
\newcommand{\lom}[1]{{\lambar\over2m_{#1}}}
\newcommand{\pol}{{\varepsilon}}
\newcommand{\pols}{{\varepsilon^*}}
\newcommand{\xqcd}{X_{\mbox{\scriptsize QCD}}}
\newcommand{\lambar}{\overline{\Lambda}}
\renewcommand{\cite}[1]{[\citen{#1}]}

\newcommand{\subsecref}[2]{Section~\protect\ref{#2}}

\begin{document}

\begin{titlepage}
{\large
\hspace*{\fill}EFI-92-36\\
\hspace*{\fill}hep-ph/9209263\\
\hspace*{\fill}September 21, 1992\\
}
\vspace{1in}

\begin{center}
	{\LARGE \bf Heavy Quark Symmetry Violation in
	Semileptonic Decays of D Mesons}\\
\end{center}
\vspace{0.5in}

\begin{center}
	{\large James F. Amundson\protect\footnotemark[1] and
	Jonathan L. Rosner\\ \it \EFI}
\end{center}
\vspace{0.5in}

\begin{center}
	{\large \bf Abstract}
\end{center} \bigskip The decays of $D$ mesons to $K l \nu$ and $K^* l \nu$
final states exhibit significant deviations from the predictions of
heavy-quark symmetry, as one might expect since the strange quark's mass is
of the same order as the QCD scale. Nonetheless, in order to understand
where the most significant effects might lie for heavier systems (such as
$B \to D l\nu$ and $B \to D^* l\nu$), the pattern of these deviations is
analyzed from the standpoint of perturbative QCD and ${\cal O}(1/m_s)$
corrections. Two main effects are noted. First, the perturbative QCD
corrections lead to an overall decrease of predicted rates, which can be
understood in terms of production of excited kaonic states. Second, ${\cal
O}(1/m_s)$ effects tend to cancel the perturbative QCD corrections in the
case of $Kl\nu$ decay, while they have minimal effect in $K^*l\nu$ decay.

\footnotetext[1]{e-mail: \tt amundson@yukawa.uchicago.edu}
\end{titlepage}

\tableofcontents

\section{Introduction}

In atomic physics, one can often separate the nuclear and electronic
effects from one another. Similar progress has been made for strongly
interacting particles within the past few years \cite{VS,NW,PW,IW}.
Particles containing a single heavy quark can be thought of as analogues of
atoms for the strong interactions. The single heavy quark behaves as the
nucleus, while the light quarks and gluons behave as the electron cloud. To
a large extent the degrees of freedom of the heavy quark and those of the
remaining matter can be discussed separately. The resulting description
validates some (but not all) older quark-model results in a systematic way,
and has come to be known as heavy quark symmetry. It represents a limit of
quantum chromodynamics (QCD) for infinite mass $m_Q$ of the heavy quark
$Q$.

The semileptonic decays of mesons containing a single heavy quark are an
ideal laboratory for testing heavy quark symmetry. Both the leading-order
results and the ${\cal O}(1/m_Q)$ corrections
\cite{Lukefinitemass,FGLfinitemass} to them have been fully worked out. In
practice the symmetry is applicable only to the semileptonic decays $B \to
D l \nu$, $B \to D^* l \nu$, and possibly to decays involving excited $D$
mesons in the final state. Only the $b$ quark (in the $B$ meson) and the
charmed quark (in the $D$ and $D^*$) are heavy enough compared to the QCD
scale of several hundred MeV that one can begin to think of applying heavy
quark symmetry. It is possible to describe six form factors (two for $B \to
D l \nu$ and four for $B \to D^* l \nu$) in terms of a single universal
function $\xi(w)$ of the variable $w \equiv v \cdot v'$, where $v$ and $v'$
are the four-velocities of the initial and final states. We shall refer to
$\xi$ as the Isgur-Wise function.

One would like very much to know the size of corrections to the universal
Isgur-Wise behavior in $B$ semileptonic decays. One benefit of this
information would be the improved ability to not only measure the $b \to c$
weak coupling accurately \cite{IWVcb,NeubertVcb}, but to be able to
reliably estimate the theoretical error. However, up to now it has not been
possible to estimate the actual magnitude of the corrections.

Early estimates of semileptonic decays were not restricted to cases in
which heavy-quark symmetry was guaranteed to work. In particular, many
attempts were made to describe the semileptonic decays of $D$ mesons within
the framework of various quark models \cite{BW,AW,KS,IS,LMMS,BBD}. The main
shortcoming of these descriptions was that they predicted rates for $D \to
K^* l \nu$ which were too large. An attempt was made \cite{GS} to isolate
the source of the discrepancy by analysis of specific form factors, but the
underlying physics remained elusive.

One should not expect heavy quark symmetry to apply to $D$ meson decays,
since no quark in the final state is heavy enough compared to the QCD
scale. Nonetheless, if regarded as a constituent of hadrons, the strange
quark has an effective mass of about 1/2 GeV, which is large enough that
one might at least expect some vestiges of heavy quark symmetry to apply.
In the present paper we take this point of view. We compare the predictions
of heavy quark symmetry for the decays $D \to K l \nu$ and $D \to K^* l
\nu$ with experiment, extracting the nonleading form factors which account
for the corrections. From these we attempt to isolate the physics
responsible for the violation of the symmetry. The question of whether
heavy-quark symmetry applies in this case, and even in the case of $K_{l3}$
decays, was raised some time ago \cite{BJLaThuile,BJ}. Others \cite{Ali}
have shown that the leading-order predictions of heavy-quark symmetry are
not sufficient to explain the branching ratios in semileptonic $D$ decays.

In brief, we have found that both the perturbative QCD and ${\cal
O}(1/m_s)$ effects are important.

There is an overall suppression of decay rates to the $K l \nu$ and $K^*
l \nu$ final states due to perturbative QCD corrections
\cite{VS,PW,FalknGrinstein,Neubertas,Neubertastwo}. The ${\cal O}(1/m_s)$
corrections add together such that they tend to cancel this suppression in
the case of $Kl\nu$ decay. In the case of $K^*l\nu$ decay, however, the
${\cal O}(1/m_s)$ corrections tend to cancel {\em each other}, leaving the
net suppression from the perturbative QCD correction.

Since the total semileptonic decay rate for $D$ mesons is close to that
expected for free charmed quarks, there must be important final states
besides $K l \nu$ and $K^* l \nu$. (We recognize that a recent search for
such final states has yielded only upper limits \cite{Anjos92}, and that
one reasonably successful description of the inclusive semileptonic rates
in terms of known final states has appeared \cite{Bai91}.) We are then led
to conclude that if there are indeed ``missing'' final states, they are
produced primarily at the expense of the $K^* l \nu$ channel, which would
help us to identify their nature.

A previous analysis of $D$ semileptonic decays in the same spirit as ours has
appeared \cite{Ito}.  Our analysis is somewhat more explicit about the sources
of heavy quark symmetry breaking, and differs in conclusions.

We have organized this paper as follows. We begin by describing the
predictions of heavy quark symmetry in \secref{sec:HQ}. After a discussion
of lifetimes and branching ratios in the data and the leading-order heavy
quark theory (\secref{sec:rates}), we use $B$ decays to determine the
Isgur-Wise function $\xi$ and the $D\to K^* l \nu$ differential branching
ratio to determine the size of the perturbative QCD corrections in $D$
decays (\secref{sec:EPD}). We then apply the theory to $D$ decay form
factors in \secref{sec:FF}. The implications of this analysis for $B$
decays is discussed in \secref{sec:IBD}\@. \secref{sec:Conclusions}
concludes.

\section{Heavy Quark Symmetry}
\label{sec:HQ}
\subsection{Leading-order Results} \label{subsec:HQ-lo}

In this subsection we discuss both the formalism and leading-order results
of heavy quark symmetry. In the following subsections we will discuss both
perturbative QCD $[{\cal O}(\alpha_s)]$ and finite-mass $[{\cal
O}({1/m_q})]$ corrections.

It is most convenient to talk about heavy quark processes in terms of
velocities instead of momenta. The most general velocity-dependent form
factors for the decay of a pseudoscalar meson $H$ to a pseudoscalar (vector)
meson $h$ ($h^*$) \cite{Neubertff} are:
	\begin{eqnarray} \label{eq:xidefns}
	\bra{h(v')}V_\mu\ket{H(v)}&=&\sqrt{\mH\mh}
		\left[\xi_+(\w)(v+v')_\mu +	\xi_-(\w)(v-v')_\mu\right],
		\nonumber\\
	\bra{h^*(v',\pol)}V_\mu\ket{H(v)}&=&i\sqrt{\mH\mhstar}
		\xi_V(\w)\epsilon_{\mu\nu\alpha\beta} \pols^\nu v'^{\alpha}
		v^{\beta},\nonumber\\
	\bra{h^*(v',\pol)}A_\mu\ket{H(v)}&=&\sqrt{\mH\mhstar}
		\left[\xi_{A_1}(\w)(\w+1)\pols_\mu \right. \nonumber\\
	&&\left. - \xi_{A_2}(\w)\pols\cdot v v_\mu -
		\xi_{A_3}(\w)\pols\cdot vv'_\mu\right],
	\end{eqnarray}
where $\pol$ is the $h^*$'s polarization vector and $\w \equiv v\cdot v' =
(m^2_H+m^2_{h^{(*)}}-q^2)/(2\mH\mhh)$. Throughout this work we denote the
mass of the heavy meson $H$ ($h$) by $m_H$ ($m_h$) and the mass of the
corresponding heavy quark by $m_Q$ ($m_q$). The allowed kinematic range is
given by
	\begin{equation}
	1 \leq \w \leq \wmax,
	\end{equation}
where
	\begin{equation}
	\wmax \equiv
		{1\over2}\left(\zz+{1\over \zz}\right)
	\end{equation}
and $\zz \equiv \mhh/\mH$. \tabref{tab:kinematics} gives numerical values
of $\zz$ and $\wmax$ for $B$ and $D$ semileptonic decays.
	\renewcommand{\arraystretch}{1.2}
	\begin{table}
	\caption{Kinematic factors in heavy meson semileptonic decays.}
	\label{tab:kinematics}
	\begin{center}
		\begin{tabular}{lcc}\hline\hline
		Decay & $\zz$ & $\wmax$\\ \hline
		$B\rightarrow D e\nu$&$0.35$&$1.59$\\
		$B\rightarrow D^* e \nu$&$0.38$&$1.50$\\
		$D\rightarrow K e \nu$&$0.27$&$2.02$\\
		$D\rightarrow K^* e\nu$&$0.48$&$1.28$\\
		\hline \hline
		\end{tabular}
	\end{center}
	\end{table}

In the absence of finite-mass and radiative corrections
	\begin{equation} \label{eq:loOne}
	\xi_+ = \xi_V = \xi_{A_1} = \xi_{A_3} \equiv \xi
	\end{equation}
and
	\begin{equation} \label{eq:loTwo}
	\xi_- = \xi_{A_2} \equiv 0,
	\end{equation}
where $\xi(\w)$ is the Isgur-Wise function. This symmetry is broken by
both radiative and finite-mass corrections.

The Isgur-Wise function satisfies the zero-recoil condition
	\begin{equation} \label{eq:iwnorm}
	\xi(1) = 1.
	\end{equation}
It has become standard in the literature to characterize its behavior for
$w$ near unity by
	\begin{equation} \label{eq:rhodef}
	\iw = 1 - \rho^2(\w - 1).
	\end{equation}
Common parameterizations of the Isgur-Wise function include the monopole,
dipole, or more generally, the $n$-pole function
	\begin{equation} \label{eq:npoleff}
	\iw = {1 \over \left(1 +{\rho^2\over n}(\w - 1)\right)^n}.
	\end{equation}
Another possible form is an exponential
	\begin{equation} \label{eq:expff}
	\iw = \exp\left[-\rho^2(\w-1)\right].
	\end{equation}

Other common parameterizations \cite{Neubertff} fall in between the
exponential and monopole forms. Since \eqn{eq:expff} is the $n\rightarrow
\infty$ limit of \eqn{eq:npoleff}, we use the monopole and exponential
forms to represent a reasonable range of forms for the Isgur-Wise function.

\subsection{Perturbative QCD Corrections} \label{subsec:HQ-pQCD}

The perturbative QCD [i.\ e., ${\cal O}(\alpha_s)$] corrections to the
heavy quark limit are calculable and, as such, have been extensively
studied \cite{VS,PW,FalknGrinstein,Neubertas,Neubertastwo}. There are two
approaches to calculating the perturbative QCD corrections in the effective
theory. The two schemes involve different schemes for matching the
effective theory on to the full theory. One approach is to assume the scales
$m_q$ and $m_Q$ are well separated and then do matching both at $m_q$ and
$m_Q$ \cite{VS,PW}. A second approach is to do the matching at some
intermediate scale $\mu$ \cite{FalknGrinstein}. Hybrid approaches have also
been developed \cite{Neubertas, Neubertastwo}.

For $B \rightarrow D$ decays these corrections are well defined. For $D
\rightarrow K$ decays, however, the situation is somewhat ambiguous.
Perturbation theory makes sense at the scale $m_c$, but, whichever approach
we use, we inevitably refer to the quantity $\alpha_s(m_s)$. There are
several ambiguities in using $\alpha_s(m_s\approx 500\mbox{ MeV})$: There
is a strong dependence on the precise value of $m_s$. Even given a specific
value of $m_s$, there is a large experimental uncertainty in the precise
value of $\alpha_s(m_s)$. Finally, the value of $\alpha_s(m_s)$ is large,
so higher-order corrections can be substantial. We will have to be
satisfied with finding the rough effects of perturbative QCD corrections
for $D \rightarrow K$ decays.

In the spirit of looking for the rough effects of the perturbative QCD
corrections, we choose to use the simplest set of radiative corrections
that contain the basic physics of the process. Therefore, for this work we
use the radiative corrections given in \refref{Neubertas}. This work is a
hybrid method where the matching is done at an intermediate scale and the
leading logarithms are summed. The result is
	\begin{equation} \label{eq:LOplusQCD}
	\xi_i(\w) = \xqcd\left[c_i+{\alpha_s(\mu) \over
		\pi}\tilde{\beta}_i(\w)\right]\xi(\w),
	\end{equation}
where
	\begin{equation}
	\xqcd=Z_{IR}(\w)\left({\alpha_s(m_Q) \over
		\alpha_s(m_q)}\right)^{-6/(33-2n_f)},
	\end{equation}
and $c_i=\{1,0,1,1,0,1\}$ for $i=\{+,-,V,A_1,A_2,A_3\}$. The mass scale
$\mu$ is chosen at some intermediate scale between $m_q$ and $m_Q$. See
\refref{Neubertas} for the definitions of the functions $\tilde{\beta}_i$
and $Z_{IR}$. \ntabref{tab:XQCD}{tab:vector} give numerical values of the
relevant functions.

	\renewcommand{\arraystretch}{1.2}
	\begin{table}
	\caption{Perturbative QCD correction common to all form factors in
	$D\rightarrow K^{(*)}$ decays.}
	\label{tab:XQCD}
	\begin{center}
	\begin{tabular}{cc} \hline\hline
		$\w$ &	$\xqcd$ \\ \hline
		1.00 & 1.269 \\
		1.20 & 1.227 \\
		1.40 & 1.189 \\
		1.60 & 1.155 \\
		1.80 & 1.124 \\
		2.00 & 1.096 \\ \hline\hline
	\end{tabular}
	\end{center}
	\end{table}

	\renewcommand{\arraystretch}{1.2}
	\begin{table}
	\caption{Perturbative QCD functions for $D \to K$ decays}
	\label{tab:pseudo}
	\begin{center}
	\begin{tabular}{ccc} \hline\hline
		$\w$ &	$\tilde{\beta}_+$ &	$\tilde{\beta}_-$ \\ \hline

		1.00 &		$-0.982$ &			$-0.260$ \\
		1.20 & 		$-1.068$ & 			$-0.254$ \\
		1.40 & 		$-1.153$ &			$-0.248$ \\
		1.60 & 		$-1.238$ & 			$-0.241$ \\
		1.80 & 		$-1.320$ & 			$-0.235$ \\
		2.00 & 		$-1.401$ & 			$-0.229$ \\ \hline\hline
	\end{tabular}
	\end{center}
	\end{table}

	\renewcommand{\arraystretch}{1.2}
	\begin{table}
	\caption{Perturbative QCD functions for $D \to K^*$ decays}
	\label{tab:vector}
	\begin{center}
	\begin{tabular}{ccccc} \hline\hline
		$\w$ &	$\tilde{\beta}_V$ &	$\tilde{\beta}_{A_1}$&
		$\tilde{\beta}_{A_2}$&	$\tilde{\beta}_{A_3}$ \\ \hline

		1.00 &	 $-0.315$ &	 $-1.648$ &	 $-1.198$ &	 $-1.229$ \\
		1.05 &	 $-0.333$ &	 $-1.644$ &	 $-1.180$ &	 $-1.231$ \\
		1.10 &	 $-0.350$ &	 $-1.641$ &	 $-1.162$ &	 $-1.233$ \\
		1.15 &	 $-0.368$ &	 $-1.638$ &	 $-1.145$ &	 $-1.237$ \\
		1.20 &	 $-0.386$ &	 $-1.637$ &	 $-1.129$ &	 $-1.240$ \\
		1.25 &	 $-0.404$ &	 $-1.636$ &	 $-1.113$ &	 $-1.245$ \\
		1.30 &	 $-0.422$ &	 $-1.636$ &	 $-1.098$ &	 $-1.249$ \\ \hline\hline
	\end{tabular}
	\end{center}
	\end{table}

We are now left with the ambiguity in choosing $\mu$, the scale at which to
do the matching of the effective theory to the full theory. The depth of
this problem is illustrated in \fig{fig:alphas}, where we have plotted
$\alpha_s$ over the range in question. For the $B \rightarrow D$ case
Neubert \cite{Neubertas,Neubertastwo} has used the summation of the leading
logs as a guide for picking the appropriate scale. To do this, notice that,
under the renormalization group \cite{VS,PW,FalknGrinstein},
	\begin{equation}
	1+{\alpha_s(\mu)\over\pi}\ln {\alpha_s(m_Q)\over\alpha_s(m_q)} \to
	\left({\alpha_s(m_Q) \over \alpha_s(m_q)}\right)^{-6/(33-2n_f)}.
	\end{equation}
We can then use this to choose the scale $\mu$ such that the two
expressions are equal. Following this prescription for $D \to K$ decays
gives $\alpha_s(\mu) = 0.73$. The uncertainty in $\alpha_s(m_c)$ induces an
uncertainty of $0.13$ in this quantity. The uncertainty due to the other
effects mentioned above is hard to estimate, but it is certainly sizable.
Later we will use the data to estimate the value of $\alpha_s(\mu)$
appropriate for $D \to K$ decays. Note that the relevant term in
\eqn{eq:LOplusQCD} is $\alpha_s(\mu)/\pi$, so the corrections due to an
$\alpha_s(\mu)$ of $0.73$ are $0.73/3.14 \simeq 25\%$.

\subsection{Finite-mass Corrections} \label{subsec:HQ-finMass}
The leading-order heavy quark results given in \twoeqn{eq:loOne}{eq:loTwo}
can be easily obtained using the trace formalism
\cite{traceformalismpeople}. In this formalism
	\begin{equation} \label{eq:lotrace}
	\langle h^{(*)}(v')|\Gamma\ket{H(v)} =
		- \sqrt{\mhh\mH}\iw\tr[\overline{\bold{h^{(*)}}}\Gamma\bold{H}],
	\end{equation}
where
	\begin{equation}
	\bold{h} = - {1+\slashchar{v} \over 2}\gamma_5,
	\end{equation}
	\begin{equation}
	\bold{h^{*}} = {1+\slashchar{v} \over 2}\slashchar{\varepsilon},
	\end{equation}
	\begin{equation}
	\overline{\bold{h}} = \gamma_0\bold{h}^{\dag}\gamma_0
	\end{equation}
and $\varepsilon$ is the vector meson's polarization vector. $\bold{H}$ is
defined similarly. In meson decays the current $\Gamma$ can be either
$V_\mu$ or $A_\mu$. This formalism is the most convenient for considering
the leading-order finite mass corrections
\cite{Lukefinitemass,FGLfinitemass}. There are subleading terms of ${\cal
O}(\lambar/m_q)$ and ${\cal O}(\lambar/m_Q)$, where $Q$ ($q$) is the heavy
quark associated with with the meson $H$ ($h$). To ${\cal O}
{(1/m_{Q(q)})}$,
	\begin{equation}
	\lambar = m_{\overline{H}}-m_Q =m_{\overline{h}}-m_q.
	\end{equation}
The subscripts $H$ and $h^{(*)}$ are barred to indicated that there is no
distinction between the pseudoscalar and vector masses to this order. We
take this average mass to be the spin average, i.\ e.,
$m_{\overline{h}}=(3\mhstar+\mh)/4$. The quantity $\lambar$ is a
fundamental parameter of QCD which is not calculable in perturbation
theory. The experimental determination of this quantity is very important.
Quark models suggest $\lambar = 300$ MeV. With this choice of $\lambar$, we
have $m_b = (3m_{B^*}+m_B)/4 -\lambar = 5.02$ GeV, $m_c = (3m_{D^*}+m_D)/4
-\lambar = 1.67$ GeV and $m_s = (3m_{K^*}+m_K)/4 -\lambar = 0.49$ GeV. (A
value $\lambar = 0.5$ GeV, obtained from QCD sum rules \cite{NewNeubert},
would imply that the quark masses are lighter, and an expansion in powers
of $\lambar/2m_s$ would not be valid.) In general there are corrections to
infinite-mass limit of both order $1/m_Q$ and $1/m_q$, but we keep only the
dominant (i.\ e., $1/m_q$) terms. In this approximation \eqn{eq:lotrace}
becomes \cite{Lukefinitemass}
	\begin{eqnarray} \label{eq:sltrace}
	\lefteqn{\langle h^{(*)}(v')|\Gamma\ket{H(v)} =
		\sqrt{\mhh\mH}\iw\tr[\overline{\bold{h^{(*)}}}\Gamma\bold{H}]}
		&&\nonumber \\
	&&\hspace*{.2in}+~\sqrt{\mhh\mH}{\lambar\over2m_q}\left\{
		\psi_1(\w)\tr[\overline{\bold{h^{(*)}}}\Gamma\bold{H}]
		\right. \nonumber\\
	&&\hspace*{.2in}+~i\psi_2(\w)\tr[v_\mu\gamma_\nu
		\overline{\bold{h^{(*)}}}\sigma^{\mu\nu}
		\mbox{$1+\slashchar{v}'\over
		2$}\Gamma\bold{H}] + \psi_3(\w)\tr[\sigma_{\mu\nu}
		\overline{\bold{h^{(*)}}}\sigma^{\mu\nu}
		\mbox{$1+\slashchar{v}'\over
		2$}\Gamma\bold{H}] \\
	&&\hspace*{.2in}+~\left[\psi_+(\w)(v_\mu+v_\mu^\prime) +
		\iw(v_\mu-v_\mu^\prime)\right]
		\tr\left[\overline{\bold{h^{(*)}}}\gamma^\mu\Gamma\bold{H}\right]
		\nonumber\\
	&&\hspace*{.2in}-~\left.\left[\psi_+(\w)(1+\w)-\iw(1-\w)\right]
		\tr\left[\gamma_\mu\overline{\bold{h^{(*)}}}\gamma^\mu
		\Gamma\bold{H}\right]
		\right\}. \nonumber
	\end{eqnarray}
To this order there are four new undetermined functions $\psi_{1,2,3,+}$.
These dimensionless functions are related to the dimensionful functions
defined in \refref{Lukefinitemass} by
	\begin{equation}
	\chi_{1,2,3} = {\lambar \over 2}\psi_{1,2,3}
	\end{equation}
and
	\begin{equation}
	\xi^{\mbox{\scriptsize Luke}}_+ = {\lambar \over 2}\psi_+.
	\end{equation}
The superscript ``Luke'' is to distinguish the $\xi_+$ used in
\refref{Lukefinitemass}, which is one of the four unknown functions
introduced at this order, from our $\xi_+$, defined in \eqn{eq:xidefns}.
The functions $\psi_1$ and $\psi_3$ satisfy the constraint
\cite{Lukefinitemass}
	\begin{equation} \label{eq:zerorecoil}
	\psi_1(\w = 1) = \psi_3(\w = 1) = 0.
	\end{equation}
This constraint, which has come to be known as Luke's theorem, has been
shown to follow from the Ademollo-Gatto theorem \cite{Boyd}.

Eqs.\ (\ref{eq:lotrace}) and (\ref{eq:sltrace}) can be understood
schematically via Figs.~\ref{fig:lodiagram} and \ref{fig:sldiagram}. In
\fig{fig:lodiagram} the decay of the meson $H$ is described by the decay of
the heavy quark $Q$ via the current $\Gamma$. This is represented by the
trace in \eqn{eq:lotrace}. The light degrees of freedom (the dashed line)
factor out of the trace to become the Isgur-Wise function $\xi$. At
sub-leading order the current $\Gamma$ is modified and the heavy quark can
interact with the light degrees of freedom. The interaction of the heavy
quark $q$ with the light degrees of freedom is represented schematically by
\fig{fig:sldiagram}. The heavy quark $q$ can interact in a spin-indepent
way yielding the same trace, but a new undetermined function ($\psi_1$) for
the light degrees of freedom. There can be two different spin-dependent
interactions. The first has the spin of the heavy quark ($\sigma^{\mu\nu}$)
interacting with the light degrees of freedom and the velocity of the
parent quark, yielding the function $\psi_2$. The factor $(1 +
\slashchar{v}')/2$ represents the propagation of the heavy quark. Note that
	\begin{equation}
	{1 + \slashchar{v}'\over2}\bold{h^{(*)}} = \bold{h^{(*)}}.
	\end{equation}
The second spin-dependent interaction corresponds to the hyperfine
interaction in the quark model. It is represented by $\psi_3$. The
modification of the current $\Gamma$ produces two new trace structures, but
can be shown \cite{Lukefinitemass} to only require the combination of the
original Isgur-Wise function and one new unknown function $\psi_+$.

The net effect on the form factors is as follows \cite{finitemasstraces}:
	\begin{equation}
	\xi_+ = \xi + \lom{q}\left[\psi_1-2(\w-1)\psi_2+6\psi_3\right]
	\end{equation}
	\begin{equation}
	\xi_- = \lom{q}\left[(2-\w)\xi+(\w+1)\psi_+\right]
	\end{equation}
	\begin{equation}
	\xi_V = \xi +
	\lom{q}\left[\xi+\psi_1-2\psi_3\right]
	\end{equation}
	\begin{equation}
	\xi_{A_1} = \xi + \lom{q}\left[{\w-1\over\w+1}\xi+
	\psi_1-2\psi_3\right]
	\end{equation}
	\begin{equation}
	\xi_{A_2} =
	\lom{q}\left[-\xi+2\psi_2+\psi_+ \right]
	\end{equation}
	\begin{equation}
	\xi_{A_3} = \xi + \lom{q}\left[\psi_1 - 2\psi_2 - 2\psi_3 + \psi_+\right]
	\end{equation}

\subsection{Momentum-dependent Form Factors}
Experiments measure the momentum-dependent form factors defined by [cf.\
the velocity dependent form factors defined in \eqn{eq:xidefns}]
	\begin{eqnarray}
	\bra{h(p')}V_\mu\ket{H(p)}&=& f_+(q^2)(p+p')_\mu +
		f_-(q^2)(p-p')_\mu \nonumber \\
	\bra{h^*(p',\pol)}V_\mu\ket{H(p)}&=&i{V(q^2) \over m_H +
		m_{h^*}}\epsilon_{\mu\nu\alpha\beta} \pols^\nu v'^{\alpha}v^{\beta}
	\nonumber \\
	\bra{h^*(p',\pol)}A_\mu\ket{H(p)}&=
		&(m_H + m_{h^*})A_1(q^2)\pols_\mu - \nonumber \\
		&&{A_2(q^2) \over m_H + m_{h^*}} (\pols\cdot p)(p+p)_\mu - \nonumber \\
		&&{A_3(q^2) \over m_H + m_{h^*}} (\pols\cdot p)(p-p)_\mu.
	\end{eqnarray}

The form factors $f_-$ and $A_3$ lead to contributions proportional to the
electron mass and are thus unmeasurable. The four measurable form factors
are related to the velocity-dependent form factors defined in
\eqn{eq:xidefns} as follows:
	\begin{equation} \label{eq:fplusconversion}
	f_+ = {z+1 \over 2 \sqrt{z}}[\xi_+ + {z-1 \over z+1}\xi_-],
	\end{equation}
	\begin{equation} \label{eq:Aoneconversion}
	A_1 = {\sqrt{\mh \mH} \over \mh + \mH}(\w+1)\xi_{A_1},
	\end{equation}
	\begin{equation} \label{eq:Atwoconversion}
	A_2 = {\mh+\mH \over 2 \sqrt{\mh\mH}}[z^*\xi_{A_2} + \xi_{A_3}].
	\end{equation}
and
	\begin{equation} \label{eq:Vconversion}
	V = {\mh+\mH \over 2 \sqrt{\mh\mH}}\xi_V.
	\end{equation}

\subsection{Heavy Quark Symmetry to ${\cal O}(\alpha_s)$ and ${\cal
O}(1/m_q)$}
Considering both subleading and perturbative QCD correction terms,
\eqn{eq:fplusconversion} becomes
	\begin{equation}
	f_+ = {z+1 \over 2 \sqrt{z}}\left\{a_f\xqcd\xi + \lom{q} \left[{z-1\over
		z+1}(2-\w)\xi+\Psi_f\right]\right\},
	\end{equation}
where
	\begin{equation}
	a_f \equiv \left[1+{\alpha_s\over\pi}
		\left(\tilde{\beta}_+ + {z-1\over z+1}
		\tilde{\beta}_-\right)\right]
	\end{equation}
and
	\begin{equation} \label{eq:Psifdef}
	\Psi_f \equiv \psi_1 -
		2(\w-1)\psi_2 + 6\psi_3 + {z-1\over z+1}(\w+1)\psi_+.
	\end{equation}
Proceeding similarly with the other form factors, we obtain
	\begin{equation}
	A_1 = {\sqrt{\mhstar \mH} \over \mhstar + \mH}(\w+1)\left[a_{A_1}\xqcd\xi
		+ \lom{q}\left( {\w-1\over \w+1}\xi +\Psi_{A_1}\right)\right],
	\end{equation}
where
	\begin{equation}
	a_{A_1} \equiv \left[1+{\alpha_s\over\pi}\tilde{\beta}_{A_1}\right]
	\end{equation}
and
	\begin{equation} \label{eq:PsiA1def}
	\Psi_{A_1} \equiv \psi_1 - 2\psi_3~;
	\end{equation}
	\begin{equation}
	A_2 = {\mhstar+\mH \over 2\sqrt{\mhstar\mH}}\left[a_{A_2}\xqcd\xi +
		\lom{q}\left(  \left({\w-1\over \w+1}-z^*\right) \xi +
		\Psi_{A_2}\right)\right],
	\end{equation}
where
	\begin{equation}
	a_{A_2} \equiv \left[1+{\alpha_s\over\pi}
		\left(z^*\tilde{\beta}_{A_2} + \tilde{\beta}_{A_3}\right) \right]
	\end{equation}
and
	\begin{equation} \label{eq:PsiA2def}
	\Psi_{A_2} \equiv \psi_1 + 2(z^*-1)\psi_2 - 2\psi_3 +(z^*+1)\psi_+
	\end{equation}
and, finally,
	\begin{equation} \label{eq:totalV}
	V = {\mhstar+\mH \over 2 \sqrt{\mhstar\mH}}\left[a_{V}\xqcd\xi +
		\lom{q}\left( \xi + \Psi_{V}\right)\right],
	\end{equation}
where
	\begin{equation}
	a_{V} \equiv \left[1+{\alpha_s\over\pi} \tilde{\beta}_{V}\right]
	\end{equation}
and
	\begin{equation} \label{eq:PsiVdef}
	\Psi_{V} \equiv \psi_1 - 2\psi_3~.
	\end{equation}
The functions $\Psi_{f,A_1,A_2,V}$ represent the contribution to the form
factors due to the unknown functions $\psi_i$. Although there are four
measured quantities and four unknown functions, there is still a relation
among the form factors:
	\begin{equation}
	\Psi_{A_1} = \Psi_V.
	\label{Psiprediction}
	\end{equation}

\section{Lifetimes and branching ratios}
\label{sec:rates}

In this section we shall discuss the total $D$ meson lifetimes, the inclusive
semileptonic decay rates, and the rates for decays to exclusive final states.
We shall also compare exclusive decay rates with results of specific models
\cite{BW,AW,KS,IS,LMMS,BBD,GS}, to show that many such models
encounter problems in describing the data.

\subsection{A Brief Tour Through the Data}
The charged and neutral $D$ mesons have different lifetimes and semileptonic
branching ratios \cite{PDG}, as shown in \tabref{tab:lifetimes}.  However, the
semileptonic decay {\it rates} are compatible with one another.  Their
average is $\Gamma(D \to X e^+ \nu_e) = (1.71 \pm 0.20) \times 10^{11} {\rm s}
^{-1}$. We shall return to the last two lines of \tabref{tab:lifetimes}
presently.

\renewcommand{\arraystretch}{1.3}
\begin{table}
\caption{Lifetimes, branching ratios (\refref{PDG}), and decay rates of $D$
mesons.}
\label{tab:lifetimes}
\begin{center}
\begin{tabular}{l c c} \hline \hline
Quantity & $D^0$ & $D^+$ \\ \hline
Lifetime~$(10^{-13}$ s) & $4.20 \pm 0.08$ & $10.66 \pm 0.23$ \\
$B(D \to X e^+ \nu_e)$ (\%) & $7.7 \pm 1.2$ & $17.2 \pm 2.9$ \\
$\Gamma(D \to X e^+ \nu_e)~(10^{11}~{\rm s}^{-1})$ & $1.83 \pm 0.29$ &
   $1.61 \pm 0.27$ \\
$\Gamma(D \to [\bar K + \bar K^*]e^+ \nu_e)~(10^{11}~{\rm s}^{-1})$ &
   $1.33 \pm 0.20$ & $0.99 \pm 0.12$ \\
$[\bar K + \bar K^*]/X$ ratio & $0.73 \pm 0.15$ & $0.62 \pm 0.13$ \\
\hline \hline
\end{tabular}
\end{center}
\end{table}

The experimental data on specific semileptonic decay channels are summarized
in \tabref{tab:exclusives}.  The data we use have been summarized in
\refref{StoneCharm}.  Original sources for the branching ratios are
\refrefs{Bai91,Adler89,Anjos89,Crawford91,Wanke,Kodama91,Anjos91,Anjos90,
Adamovich91,Albrecht91,Kodama92,Alimonti}.

\renewcommand{\arraystretch}{1.2}
\begin{table}
\caption{Semileptonic branching ratios and decay rates of $D$ mesons to
$\bar K e^+ \nu_e$ (\refref{StoneCharm}) and $\bar K^* e^+ \nu_e$ .}
\label{tab:exclusives}
\begin{center}
\begin{tabular}{l c c c} \hline \hline
Decay & Ref. & Branching ratio (\%) & Rate $(10^{10} {\rm s}^{-1})$ \\ \hline
$D^0 \to K^-e^+\nu_e$ & Mark III \cite{Adler89} & $3.4 \pm 0.5 \pm 0.4$ & \\
 & E691 \cite{Anjos89} & $3.8 \pm 0.5 \pm 0.6$ & \\
 & CLEO \cite{Crawford91} & $3.8 \pm 0.3 \pm 0.6$ & \\
 & ARGUS \cite{Wanke} & $3.9 \pm 0.2 \pm 0.7$ & \\
 & E653 \cite{Kodama91} & $2.5 \pm 0.4 \pm 0.5$ & \\
 & Average  & $3.3 \pm 0.4^{a)}$         & $7.9 \pm 1.0$ \\
\hline
$D^+ \to \bar K^0 e^+ \nu_e$ & Mark III \cite{Bai91}& $6.5 \pm 1.6 \pm 0.7$ &
\\
 & E691 \cite{Anjos91} & $6.1 \pm 0.9 \pm 1.6$ & \\
 & Average  & $6.3 \pm 1.3$         & $5.8 \pm 1.2$ \\
\hline
$D \to \bar K e^+ \nu_e$ & Average &                    & $7.0 \pm 0.8$ \\
\hline
$D^0 \to K^{*-} e^+ \nu_e$ & Mark III \cite{Bai91} & $3.5 \pm 1.5^{b)}$ & \\
 & CLEO \cite{Crawford91} & $1.7 \pm 0.8^{c)}$    & \\
 & Average & $2.2 \pm 0.7$       & $5.2 \pm 1.7$ \\
\hline
$D^+ \to \bar K^{*0} e^+ \nu_e$ & Mark III \cite{Bai91}& $4.2 \pm 1.6^{d)}$ &
\\
 & E691 \cite{Anjos90}& $4.4 \pm 0.4 \pm 0.8$ & \\
 & WA82 \cite{Adamovich91} & $5.6 \pm 1.6 \pm 0.9$ & \\
 & ARGUS \cite{Wanke} & $4.2 \pm 0.6 \pm 1.0$ & \\
 & E653$^{e)}$ \cite{Kodama92} & $3.25 \pm 0.71 \pm 0.75$ & \\
 & Average & $4.1 \pm 0.5$       & $3.9 \pm 0.5$ \\ \hline
$D \to \bar K^* e^+ \nu_e$ & Average &                  & $4.0 \pm 0.5$ \\
\hline \hline
\end{tabular}
\end{center}

\leftline{$^{a)}$We became aware of the measurement of
$BR(D^0\to K^-\mu^+\nu_\mu) = 4.0\pm0.8\pm0.8\%$ \cite{Alimonti}}
\leftline{~~~after the completion of this work. This value is not included
in the average.}
\leftline{$^{b)}$Based on quoted $B(D \to [\bar K \pi] e^+ \nu_e) =
(4.4^{+1.9}_{-1.0} \pm 0.6)$\%}
\leftline{~~~multiplied by resonant fraction $0.79^{+0.15+0.09}_{-0.17-0.03}$.}

\leftline{$^{c)}$Based on quoted ratio $B(D^0 \to K^{*-} e^+ \nu_e)/
B(D^0 \to K^- e^+ \nu_e)$}
\leftline{$~~~= 0.51 \pm 0.18 \pm 0.06$
times our average for $B(D^0 \to K^- e^+ \nu_e)$.}

\leftline{$^{d)}$Based on quoted $B(D \to [\bar K \pi] e^+ \nu_e) =
(5.3^{+1.9}_{-1.1} \pm 0.6)$\%}
\leftline{~~~multiplied by resonant fraction $0.79^{+0.15+0.09}_{-0.17-0.03}$.}

\leftline{$^{e)}$Based on decay $D^+ \to \bar K^{*0} \mu^+ \nu_\mu$.}

\end{table}

The decay rates to exclusive final states for neutral and charged $D$ mesons
are consistent with one another.  The averages over charged and neutral decays
for each channel are also quoted.  If we sum over decays to $\bar K e^+ \nu_e$
and $\bar K^* e^+ \nu_e$, we obtain the rates shown in the fourth row of
\tabref{tab:lifetimes}.  These rates fall short of the total semileptonic
decay rates, though not with overwhelming statistical significance, as
shown in the last row of \tabref{tab:lifetimes}.  The average of the two
numbers in the last row of \tabref{tab:lifetimes} is $0.67 \pm 0.10$. The
Cabibbo-suppressed exclusive final states should account for about 5\% $\times$
(relative phase space) $\approx 7\%$ of the shortfall, and the nonresonant
fraction of the decay mode $D \to [\bar K \pi] e^+ \nu_e$ should account for
another $\approx 6\%$.  This still leaves about $(20 \pm 10)\%$ of the
inclusive semileptonic decays unaccounted for.

The ratio $\Gamma(D \to \bar K^* e^+ \nu_e)/\Gamma(D \to \bar K e^+ \nu_e)$ is
measured to be considerably smaller than that predicted in most models.
Experimental values for this ratio are collected in \tabref{tab:ratios}.

\begin{table}
\caption{Ratios $B(D \to \bar K^* e^+ \nu_e)/B(D \to \bar K e^+ \nu_e)$.
{}From \refref{StoneCharm}.}
\label{tab:ratios}
\begin{center}
\begin{tabular}{l c} \hline \hline
Ref. & Ratio \\ \hline
Mark III \cite{Bai91}  & $0.80^{+0.36}_{-0.26}$   \\
CLEO \cite{Crawford91} & $0.51 \pm 0.18 \pm 0.06$ \\
ARGUS \cite{Wanke}     & $0.55 \pm 0.08 \pm 0.10$ \\
E691 \cite{Anjos91}    & $0.55 \pm 0.14$ \\
Average                & $0.57 \pm 0.08$ \\
\hline \hline
\end{tabular}
\end{center}
\end{table}

The predictions of a number of models for the ratio in \tabref{tab:ratios}
are reviewed in \refref{StoneCharm}.  These predictions are typically 0.9
or higher \cite{BW,AW,KS,IS} except for the more recent lattice calculations
\cite{LMMS,BBD}, which are still quite uncertain.  One can see a residue of
the heavy-quark limit in these predictions.  In the limit of very heavy initial
and final quarks, a counting of spin degrees of freedom would then lead to a $V
l \nu/P l \nu$ ratio of 3, where $V$ and $P$ stand for the ground state vector
and pseudoscalar mesons \cite{VS}. This is not far from the actual situation in
the decays $B \to D^* l \nu$ and $B \to D l \nu$.

If the recoil of the final quark could be neglected, {\it only} the $P$ and $V$
mesons would be produced in the final state, so that $P$ would account for 1/4
and $V$ for 3/4 of the hadrons in the semileptonic final state \cite{VS}.  By
comparing the average inclusive semileptonic decay rate $\Gamma(D \to X e^+
\nu_e) = (1.71 \pm 0.20) \times 10^{11} {\rm s} ^{-1}$ with the averages in
\tabref{tab:exclusives}, we see that $K$ accounts for $(7.0 \pm 0.8)/(17.1 \pm
2.0) = 0.41 \pm 0.07$ of the $D$ semileptonic decay rate, while $K^*$ accounts
for $(4.0 \pm 0.5)/(17.1 \pm 2.0) = 0.23 \pm 0.04$ of that rate.  The
discrepancy with respect to the heavy-quark limit thus is much more marked for
the $K^*$ than for the $K$.

Our task is then to account for departures from the heavy-quark limit which
lead to a modest enhancement of the rate for $D \to K l \nu$ and a substantial
suppression of that for $D \to K^* l \nu$.  We have mentioned in the
Introduction that a combination of two effects, perturbative QCD and
${\cal O}(1/m_s)$, seems to be responsible.  Our subsequent discussion
attempts to put this claim on a quantitative footing.

Additional discrepancies with respect to the heavy-quark limit in
charmed meson decays show up in the behavior of individual form factors
\cite{GS}.  These are discussed in \secref{sec:FF}.

\subsection{Comparison with Leading-order Heavy Quark Theory}

It is convenient to have expressions for the widths in terms of the
velocity-dependent form factors. Defining
	\begin{equation}
	\Gamma_0 = {G^2_{F}|V_{Qq}|^2m_H^5 {z^{(*)}}^3\over 48\pi^3},
	\end{equation}
we find the following expressions for the branching ratios:
	\begin{equation} \label{eq:pseudorate}
	{\md{\Gamma(H\rightarrow h l \nu)}
	\over \md{\w}} =
	\Gamma_0 (z+1)^2(\w^2-1)^{3/2}\left[\xi_+(\w) +
		{z-1 \over z+1}\xi_-(\w)\right]^2.
	\end{equation}

	\begin{eqnarray} \label{eq:transrate}
	{\md{\Gamma_T(H\rightarrow h^* l \nu)} \over \md{\w}}&=&
	2\Gamma_0 (1+{z^*}^2-2{z^*}\w)(\w+1)
	(\w^2-1)^{1/2}\times \nonumber \\
	&&\left[ (\w-1)\xi_V^2(\w) +
	(\w+1)\xi_{A_1}^2(\w)\right].
	\end{eqnarray}

	\begin{eqnarray} \label{eq:longrate}
	{\md{\Gamma_L(H\rightarrow h^* l \nu)} \over \md{\w}}&=&
		\Gamma_0 (\w+1)^2(\w^2-1)^{1/2}\times \nonumber \\
	&&\left\{(\w-{z^*})\xi_{A_1}(\w) -
		(\w-1)\left[{z^*}\xi_{A_2}(\w)+
		\xi_{A_3}(\w)\right]\right\}^2.
	\end{eqnarray}
These expressions agree with those found by Neubert in \refref{NeubertVcb}.

In order to illustrate the problems encountered by heavy quark symmetry for
$D$ semileptonic decays, we adapt the above relations for the decays $D \to
\bar K e^+ \nu_e$ and $D \to \bar K^* e^+ \nu_e$. We use the relations
(\ref{eq:loOne}) and (\ref{eq:loTwo}) of \subsecref{sec:HQ}{subsec:HQ-lo},
and predict rates as a function of $\alpha_s(\mu)$ as described in
\subsecref{sec:HQ}{subsec:HQ-pQCD}. The results are shown in \fig{fig:DBR}.

We see that in the lowest order of the heavy-quark theory, in the absence
of perturbative QCD corrections, the rate for the decay $D \to \bar K e^+
\nu_e$ is not badly predicted, but that for $D \to \bar K^* e^+ \nu_e$ is
about a factor of two above the experimental value. The $K^*/K$ ratio is
predicted to lie in the range of 1.2 to 1.7 for the monopole form factor
and 1.4 to 2.2 for the exponential form factor, in contrast to the
experimental value of $0.57 \pm 0.08$. Perturbative QCD corrections lead to
an overall decrease in the predicted rates and cannot account for the
$K^*/K$ ratio. As we shall see, spin-dependent ${\cal O}(1/m_s)$
corrections are needed for that purpose.

\section{Extracting Parameters from the Data} \label{sec:EPD}
\newcommand{\boldB}{\protect\boldmath$\overline{B}$}

\subsection{Determining the Isgur-Wise Function from \boldB~Decays.}
\label{subsec:EPD-getrho}

We use both total and differential branching ratios in $B$ decay to
determine $\iw$. We have used the expressions
(\ref{eq:pseudorate})--(\ref{eq:longrate}) with $H = \overline{B}^0$; $h,h^* =
D^+, D^{*+}$; and $|V_{Qq}| = |V_{cb}|= 0.041$ \cite{Stone} with the
leading-order plus perturbative QCD form factors in \eqn{eq:LOplusQCD} in
comparison with the total branching ratios given in \refref{Stone} to
obtain the results shown in \tabref{tab:rho}.
	\begin{table}
	\caption{$\rho$ values obtained from various fits.}
	\label{tab:rho}
	\begin{center}
	\begin{tabular}{lcc} \hline\hline
		&\multicolumn{2}{c}{Fitted value of $\rho$}\\
		Decay & Monopole & Exponential\\ \hline
		\multicolumn{3}{c}{\em Integrated Branching Ratios}\\
		$B\rightarrow D e \nu$&$1.02^{+0.22}_{-0.19}$&$0.93\pm0.16$\\
		$B\rightarrow D^* e \nu$&$0.97\pm0.21$&$0.91\pm0.17$\\
		\multicolumn{3}{c}{\em Differential Branching Ratio}\\
		$B\rightarrow D^* e\nu$&$1.11\pm0.50$&$1.02^{+0.36}_{-0.42}$\\
		\multicolumn{3}{c}{\em Average}\\
		&$1.00\pm0.15$&$0.93\pm0.10$\\ \hline \hline
	\end{tabular}
	\end{center}
	\end{table}

We can also use the $B\rightarrow D^*l\nu$ differential branching ratio to
determine $\rho$. This process has been considered in detail by Neubert
\cite{NeubertVcb}. Our fit to the ARGUS data \cite{ARGUS} is shown in
\fig{fig:ARGUS}. As can be seen in \tabref{tab:rho}, our determination of
$\rho$ is dominated by the integrated branching ratios. The plot of the
differential branching ratio serves as a visual check of our results. We
have chosen the ARGUS data as representative; other data exist
\cite{BortnStone}. The Isgur-Wise function can be easily extracted from the
data by plotting the square root of the differential branching ratio
divided by the the factors that multiply the Isgur-Wise function in the
symmetry limit. This procedure can be improved by also dividing out the
perturbative QCD corrections to the $\xi_{A_1}$ piece, which dominates the
rate.

We then plot $|V_{cb}|\left[\left({1/ G}\right)\;
{\md{\Gamma}/\md{\w}}\right]^{1/2}$, where
	\begin{equation}
	G \equiv |V_{Qq}|^2{G^2_F \over 48\pi^3}
		m^5_H {z^*}^3 \sqrt{\w^2-1} (\w+1)^2 \xqcd^2 a_{A_1}^2
		\left[{4\w\over\w+1}(1-2\w z^*+{z^*}^2)+(1-z^*)^2\right] ,
	\end{equation}
which amounts to taking the $m_{Q,q}\rightarrow\infty$ and
$\alpha_s\rightarrow 0$ limit of \twoeqn{eq:transrate}{eq:longrate} and
then subtracting the perturbative QCD correction to the dominant part of
the rate. At this level of approximation, $\left[\left({1/
G}\right)\;{\md{\Gamma}/\md{\w}}\right] = \xi(w)^2$. Furthermore, the zero
recoil condition
	\begin{equation} \label{eq:normpoint}
	\left.\left({1\over G}\;{\md{\Gamma}\over\md{\w}}\right)
	\right|_{w=1} = 1
	\end{equation}
holds including corrections of ${\cal O}(\alpha_s)$ {\em and} ${\cal
O}(1/m_q)$.

In \refrefs{NeubertVcb,ARGUS} the quantity $\left[\left({1/G}\right) \;
{\md{\Gamma}/\md{\w}}\right]^{1/2}$ was labeled simply $\xi(w)$, but we
prefer the present notation, as it makes the assumptions involved explicit.
Instead of trying to fit for $|V_{cb}|$, we have used a fixed value (0.041)
because we are only interested in the overall shape of the form factor. In
\refref{NeubertVcb}, values of $\rho=1.14\pm0.23$ and $\rho=1.07\pm0.22$
were obtained using the pole form $\iw = [2/(\w+1)]^{2\rho^2}$ and the
exponential form \eqn{eq:expff}, respectively. A monopole form
(\eqn{eq:npoleff} with $n=1$) was used in \refref{Rosner} to obtain $\rho =
1.26 \pm 0.19$.

Both methods show a systematic dependence of the fitted value for $\rho$
with the functional form. \fig{fig:ARGUS} shows that the two fitted
functions look very similar with the different values of $\rho$. The
precise value of $\rho$ is important for determining $|V_{cb}|$ by
extrapolating to $\w = 1$. We are not attempting to do this here, however.
The functional dependence will have some impact if we want to extrapolate
beyond the end of the data ($\w \sim 1.4$). This is the case for $D
\rightarrow Kl\nu$ decays, where $\w_{\mbox{\scriptsize max}} = 2.0$. To
minimize the systematic error from the functional forms, we will use $\rho
= 1.0$ with the monopole form and $\rho = 0.93$ with the exponential form.

Analyticity arguments suggest that the Isgur-Wise function should have a
radius of convergence of approximately unity, i.\ e., it should have a pole
near $\w = -1$. This corresponds to a threshold for particle production
near $q^2 = (m_b + m_c)^2$. The $n$-pole function has a pole at $\w =
1-n/\rho^2$. For $n=1$ and $\rho\sim 1$ the pole is near $\w = 0$. The
monopole description can only be valid if there are a series of poles at
$\w < -1$ which somehow conspire to mimic a monopole description with a
pole near zero. This seems unlikely, so we conclude the dependence on the
functional form of the Isgur-Wise function is even less than the difference
between the monopole and exponential forms. In the remaining plots we use
the exponential form with $\rho = 0.93$.

\subsection{Determining $\alpha_s(\mu)$ from \protect\boldmath$D$ decays}
\label{subsec:EPD-getalpha}

The condition in \eqn{eq:normpoint} has been used to determine the unknown
parameter $|V_{cb}|$ from $B$ decays. Although the corresponding parameter
in $D$ decays, $|V_{cs}|$, is well known {\em a priori}, we have argued in
\subsecref{sec:HQ}{subsec:HQ-pQCD} that the appropriate value of
$\alpha_s(\mu)$ is not. In this section we use the $D \to K^*l\nu$
differential width extrapolated to the point $w=1$ to determine
$\alpha_s(\mu)$. Not only do $1/m_s$ corrections vanish at this point, but
the QCD corrections to the dominant $\xi_{A_1}$ form factor are the largest
of any of the QCD corrections, as can be seen from comparing the values of
$\tilde{\beta}_{A_1}$ with the other $\tilde{\beta}_i$'s in Tables
\ref{tab:pseudo} and \ref{tab:vector}.

The published data \cite{Kodama92,esixninekstar,esixfivekstar} includes
only raw data, i.\ e., including detector efficiencies, compared to Monte
Carlo simulations used to determine the form factors. We have turned this
around to determine a net efficiency at each point and then determine the
net distributions. The experimentalists themselves could do a much better
job of this, but our method will have to suffice until they publish their
own differential branching ratios.

In \fig{fig:getalpha} we have plotted the average of the two available
data sets along with our fit used to extrapolate to $w=1$. The fitted
function is an exponential. The data is very flat, so the dependence of
the extrapolation on the functional form is slight. The fit yields
	\begin{equation}
	\left.\left(a_{A_1}{1\over G}\;{\md{\Gamma}\over\md{\w}}\right)
	\right|_{w=1} = 0.457\pm0.024,
	\end{equation}
which gives $\alpha_s(\mu) = 1.03\pm0.11$. This value is considerably
larger than the guess we had in \subsecref{sec:HQ}{subsec:HQ-pQCD}, but the
uncertainties in that guess are very large. To get a consistent picture of
$D$ decays using only first order symmetry breaking corrections we have to
accept a large value of $\alpha_s(\mu)$. The corrections are then ${\cal
O}(\alpha_s(\mu)/\pi) \approx 33\%$.

\section{Form Factors} \label{sec:FF}

\subsection{Comparison with Experiment}
All four $D \to K$ form factors have been measured. Three groups
\cite{Adler89,Anjos89,Crawford91} have measured the $q^2$ dependence
of $f_+$. They find that the data are consistent with the monopole vector
dominance form
	\begin{equation}
	f_+(q^2) = {f_+(0) \over 1-q^2/m^2_f}.
	\end{equation}
We use the world average value given by Stone \cite{StoneCharm}
	\begin{equation}
	m_f = 2.0^{+0.3}_{-0.2}.
	\end{equation}
The normalization constant can be obtained from integrating the branching
ratio. Here, too, we use the average value cited by Stone
\cite{StoneCharm}:
	\begin{equation}
	f_+(0) = 0.69\pm0.04.
	\end{equation}
The three measurable form factors in $D\rightarrow K^*l\nu$ semileptonic
decay have also been measured \cite{Kodama92,esixninekstar,esixfivekstar}.
Unfortunately, these measurements {\em assume} the $q^2$ dependences follow
single-pole vector dominance so that
	\begin{equation} \label{eq:aexpform}
	A_{1,2}(q^2) = {A_{1,2}(0) \over 1-q^2/m^2_A}
	\end{equation}
and
	\begin{equation} \label{eq:vexpform}
	V(q^2) = {V(0) \over 1-q^2/m^2_V},
	\end{equation}
where $m_A$ and $m_V$ are assumed to be the lowest mass $c\overline{s}$
states with $J^P=1^+$ and $1^-$, i.\ e., $m_A = 2.5$ GeV and $m_V = 2.1$
GeV. The experimentalists then fit for the ratios $A_2(0)/A_1(0)$ and
$V(0)/A_1(0)$. $A_2(0)$, $A_1(0)$ and $V(0)$ then may be determined from
the total branching ratio, which is dominated by $A_1(0)$.

	\begin{table}
	\caption{Form factor measurements for $D\rightarrow K^*l\nu$.}
	\label{tab:ffexp}
	\begin{center}
		\begin{tabular}{lccccc}\hline\hline
		Experiment & $A_2(0)/A_1(0)$ & $V(0)/A_1(0)$ & $A_1(0)$ & $A_2(0)$ &
			$V(0)$\\ \hline
		E691 & $0.0\pm0.5\pm0.2$ & $2.0\pm0.6\pm0.3$ & $0.46\pm0.07$ &
			 $0.0\pm0.2$ & $0.9\pm0.3$\\
		E653 & $0.82\pm0.23$ & $2.0\pm0.3\pm0.2$ &$0.49 \pm 0.07$ &
			$0.40\pm0.14$ & $0.99\pm0.27$\\
		Average & & & $0.48\pm0.05$ & $0.27\pm0.11$ & $0.95\pm0.20$\\
		 \hline\hline
		\end{tabular}
	\end{center}
	\end{table}

\tabref{tab:ffexp} displays the values obtained by the two experiments. The
measurements of $V(0)$ and $A_1(0)$ are quite consistent with one another.
The values of $A_2(0)$ differ from one another by slightly less than two
standard deviations.

It should be pointed out that there is some confusion stemming from the way
these values are reported. The numbers $V(0)$, $A_1(0)$ and $A_2(0)$ have
been interpreted as measurements of the these form factors at a single
point ($q^2 = 0$). In fact, both experiments do their fits to
\eqn{eq:aexpform} and \eqn{eq:vexpform} {\em over the entire range of
$q^2$.} $V(0)$, $A_1(0)$ and $A_2(0)$ are merely the constants in
\eqn{eq:aexpform} and \eqn{eq:vexpform}.

In \fig{fig:ff} we compare the measured form factors with our predictions
from heavy quark symmetry. We see that the deviations from the leading-order
heavy quark predictions are about as large as was expected: $\sim
\lambar/2m_s \sim 30\%$. The rates depend on the square of the form
factors, however, so a deviation in the form factor of $25\%$ can lead to a
difference in rates by a factor of 2! This helps explain the discrepancies
seen in \secref{sec:rates}. Note that, although the $D\to Kl\nu$ rate was
closest to the leading-order prediction of heavy quark symmetry, it has
the largest deviation in the form factors. The $1/m_s$ effects cancel the
$\alpha_s$ effects for this decay, while they add coherently in $D\to
K^*l\nu$ decay.

\subsection{Determining \protect$\lambar$ from $V(w=1)$}
\label{subsec:FF-getlambar}

It follows from Eqs.~(\ref{eq:iwnorm}), (\ref{eq:zerorecoil}) and
(\ref{eq:totalV}) that
	\begin{equation} \label{eq:vone}
	V(w=1) = {\mhstar+\mH \over 2 \sqrt{\mhstar\mH}}\left(a_{V}\xqcd
	+\lom{q}\right).
	\end{equation}
A measurement of $V(w=1)$ is the theoretically cleanest way to measure the
fundamental quantity $\lambar$. Even in $D$ decays, where the size of
$\alpha_s(\mu)$ is in question, this relation is useful because the
dependence on $\alpha_s(\mu)$ is relatively weak. Using the measured $V$
from the previous section, this leads to
	\begin{equation}
	\lom{s} = -0.05\pm0.239,
	\end{equation}
and
	\begin{equation}
	\lambar = -44\pm297\mbox{ MeV}.
	\end{equation}

This error is clearly too large to be truly useful. We have the further
caveat that the extrapolation of $V(w)$ to $w=1$ will only be reliable once
the $w$-dependence of the form factors is measured. We hope this quantity
will be measured with smaller uncertainties in the future.

The above measurement of $\lambar$ is subject to the various uncertainties
involved in applying heavy quark symmetry to the strange quark. It is much
more important to measure $\lambar/(2 m_c)$, where heavy quark symmetry is
on very sound theoretical ground. CLEO \cite{CLEOBff} has recently
published a first measurement of $V(w=1)$ in $B\to D^*l\nu$ decays. They
obtain two different values from two different fits:
	\begin{equation}
	V(w=1) = \left\{
			\begin{array}{c}
			0.91\pm0.49\pm0.12\mbox{~(fit a)}\\
			1.19\pm0.57\pm0.15\mbox{~(fit b)}
			\end{array}
			\right.,
	\end{equation}
which we average to get $V(w=1) = 1.05 \pm 0.5$. Using $m_H = 5.28$ GeV,
$m_{h^*} = 2.01$ GeV, $\xqcd = 1.11$ and $a_V = 0.97$ in \eqn{eq:vone}
leads to
	\begin{equation}
	\lom{c} = -0.14\pm0.45.
	\end{equation}
The uncertainties are so large that it doesn't make sense to extract
$\lambar$. It is important that this quantity become better measured in the
future.

\subsection{Extracting the Subleading Effects From the Data}

In \fig{fig:Psi} we have assumed that the differences between the
predictions and the experiments shown in \fig{fig:ff} are completely due to
subleading terms. We have then subtracted out the $\xi$ dependent effects
to obtain $\Psi_{f,A_1,A_2,V}$. This procedure depends on the value of
$\lambar$, so we have included results for both $\lambar = 200 \mbox{ and
} 400$ MeV. The size of the experimental uncertainties depends on
$\lambar$ because we have scaled out $\lambar/(2m_s)$.

Several things are apparent from \fig{fig:Psi}: First, the dimensionless
functions all come out to be ${\cal O}(1).$ We also see that the prediction
$\Psi_{A_1} = \Psi_V$ is better satisfied for the smaller value of
$\lambar$. This is consistent with the results in
\subsecref{sec:FF}{subsec:FF-getlambar}. We also see that the corrections
to the pseudoscalar channel embodied by $\Psi_f$ are substantial and
positive. The net effect of the corrections to the vector decays is small
and negative.

For concreteness, we can attempt to estimate the individual functions
$\psi_i$ under certain assumptions. First, we choose the smaller value of
$\lambar$ in \fig{fig:Psi}. Using the definition of $\Psi_f$
[\eqn{eq:Psifdef}] and $\psi_1(1) = \psi_3(1) = 0$ (Luke's theorem), we
obtain
	\begin{equation}
	\Psi_f(1) = -1.15 \psi_+(1)
	\end{equation}
Comparing with the light band in the plot of $\Psi_f$ in \fig{fig:Psi},
we obtain $\psi_+(1) \approx -1.5$. Similarly, we use
	\begin{equation}
	\Psi_{A_2}(1) = -1.04 \psi_2 + 1.48 \psi_+
	\end{equation}
to obtain $\psi_2(1) \approx -2$. To get information about $\psi_1$ and
$\psi_3$ we need to go to another kinematical point. It is best to use the
{\em minimum} recoil point of the pseudoscalar decay, since it has the larger
kinematic range and the kinematic dependence of the the pseudoscalar form
factor ($f$) has been measured. Using \eqn{eq:Psifdef} once again, we have
	\begin{equation}
	\Psi_f(2) = \psi_1(2) - 2\psi_2(2) + 6\psi_3(2) - 1.7\psi_+(2).
	\end{equation}
Since $\Psi_{A_1}$ is consistent with zero, we set $\psi_1(w) =
2\psi_3(w)$. If we also assume $\psi_+(w)$ and $\psi_2(w)$ are roughly
constant in $w$, we can then obtain $\psi_3(2) \approx -0.5$ and
$\psi_1(2) \approx -1$. These estimates are very rough, owing to both
experimental uncertainties and our assumptions.

\section{Implications for B Decays} \label{sec:IBD}

The overall QCD decrease of semileptonic decay rates into pseudoscalar and
vector mesons found for $D$ meson decays implies that for $B$ meson decays
similar, but slightly weaker, effects are to be expected. In view of the
ambiguity reflected in \fig{fig:Psi}, we can make only qualitative
statements.

While the ${\cal O}(1/m_q)$ effects discussed above are expected to reduce
the ratio $\Gamma(B \to D^* l \nu)/\Gamma(B \to D l \nu)$ from its
infinite-mass limit \cite{VS} of 3, the lowest-order prediction of this
ratio exceeds 3 slightly for reasonable slopes of the Isgur-Wise function
\cite{Rosner}. The experimental branching ratios \cite{Stone} are $B(B \to
D^* l \nu) = (4.4 \pm 0.3 \pm 0.6)\%$ and $B(B \to D l \nu) = (1.6 \pm 0.3
\pm 0.2)\%$; their ratio is $2.75 \pm 0.75$. The error is too large to
reflect the presence of any of the effects mentioned here.

The sum of the semileptonic $B$ branching ratios to $D$ and $D^*$, $(6.0
\pm 0.8)\%$, is only 1/2 to 2/3 of the inclusive semileptonic branching
ratio \cite{Stone}, $B(B \to X l \nu) \approx 11\%$. The remaining 1/2 to
1/3 of the semileptonic decays are expected to involve excited $D$ mesons
or extra pions. A non-trivial Isgur-Wise function (one with $\rho >0$) is
sufficient to reflect the presence of such excitations \cite{BJ}; one does
not require ${\cal O}(1/m)$ effects to understand them.

Perhaps the most pressing question about heavy quark symmetry and $B$
decays is how much we can learn about $V_{cb}$. Our analysis suggests that
measuring $V_{cb}$ by extrapolating the $B\to Dl\nu$ spectrum to the
normalization point would be unreliable because of the substantial
$\lambar/m_c$ effect at that point. Fortunately, extrapolating the $B\to
D^*l\nu$ spectrum, as was done in \refref{NeubertVcb}, avoids all
complications due to $\lambar/m_c$ effects. The question in this case,
then, is the size of the $(\lambar/m_c)^2$ corrections. We have shown in
\subsecref{sec:FF}{subsec:FF-getlambar} how the quantity $\lambar/(2m_c)$
can be extracted from the data. Although the error we obtained is very
large, the results favor a small value of $\lambar$, which would be good
news for an accurate determination of $V_{cb}$ if more accurate
measurements confirm that $\lambar$ is indeed small.

\section{Conclusions} \label{sec:Conclusions}

We have applied a broken version of heavy-quark symmetry to the
semileptonic decays $D \to K l \nu \mbox{ and }D \to K^* l \nu$, in order
to study the deviations from the symmetry in an environment where they are
particularly prominent. The price that we pay is the possible loss of
validity of the symmetry. We have to regard the strange quark as heavy,
which we can do if we regard it as having a constituent-quark mass of order
1/2 GeV.

We have been able to identify several physical sources of the apparent
deviations from heavy-quark symmetry (and of the shortcomings of early
models for $D$ semileptonic decays).

First, perturbative QCD effects lead to an overall reduction in predicted
$D \to K l \nu\mbox{ and }D \to K^* l \nu$ rates with respect to the
lowest-order theory. One expects the remainder of the semileptonic decays
to show up in excited kaons or $K + n \pi$ final states. No evidence for
these final states exists at present.

Second, ${\cal O}(1/m_s)$ effects give different contributions to the
pseudoscalar and vector final states. These effects add to cancel the
perturbative QCD effects in the pseudoscalar final state. In the vector
final state, however, these effects cancel one another, leaving the overall
supression due to the perturbative QCD effects untouched. This explains
the suppression of the vector final state relative to the pseudoscalar
final state.

We find that it is instructive to view the strange quark, under some
circumstances, as ``heavy,'' which allows us to gain a qualitative
understanding of effects in $D$ decays which are still too small to show
up in the decays $B\to Dl\nu$ and $B\to D^* l\nu$. A study of these
decays with increased precision would provide an excellent check of our
conclusions.

\section*{Acknowledgments}

We would like to thank G. Boyd, N. Isgur, M. Luke, M. Neubert, S. Stone and
M. Wise for helpful discussions. One of us (JFA) would like to thank the
Institute for Theoretical Physics in Santa Barbara, where part of this work
was completed.

\renewcommand{\textfraction}{0}
\renewcommand{\topfraction}{1}
\renewcommand{\floatpagefraction}{0}
\setcounter{totalnumber}{99}

\newpage
\section*{Figure Captions}

\begin{figure}[h]
\caption{Variation of $\alpha_s$ over the region of interest for $D$
decays based on \protect\cite{Kwong} $\alpha_s(m_c)~=~0.276 \pm 0.014$
The solid line represents the central value, while the dashed lines
represent the one standard deviation uncertainties.}
\label{fig:alphas}
\end{figure}

\begin{figure}[h]
\caption{Diagram for heavy meson decays at leading order.~ ~ ~ ~ ~ ~ ~
{} ~ ~ ~ ~ ~ ~ ~ ~ ~ ~ ~ ~ }
\label{fig:lodiagram}
\end{figure}

\begin{figure}[h]
\caption{Subleading diagram where the heavy quark interacts with the
light degrees of freedom.}
\label{fig:sldiagram}
\end{figure}

\begin{figure}[h]
\caption{Predictions of heavy-quark symmetry for decay rates of $D \to
\bar K e^+ \nu_e$ (a,c) and $D \to \bar K^* e^+ \nu_e$ (b,d).
Predictions without QCD corrections (i.\ e., $\alpha_s(\mu) = 0$ and
$\protect\xqcd = 1$) are shown in (a,b) as functions of $\rho$ for
monopole (solid lines) and exponential (dashed lines) form factors. The
shaded bands correspond to the experimental values with one standard
deviation uncertainties.
The ranges of $\rho$ allowed by the fits to $B$ decays in
\protect\subsecref{sec:EPD}{subsec:EPD-getrho} are shown for monopole
(circles) and exponential (diamonds) form factors. Predictions for
central values of $\rho$ are shown as functions of $\alpha_s(\mu)$
(holding $\protect\xqcd$ fixed) in (c,d), where the bars indicate the
range of $\alpha_s(\mu)$ obtained in
\protect\subsecref{sec:EPD}{subsec:EPD-getalpha}.}
\label{fig:DBR}
\end{figure}

\begin{figure}[h]
\caption{Data taken from \protect\refref{ARGUS}. The solid and dashed
lines are fits to the exponential and monopole forms, respectively.}
\label{fig:ARGUS}
\end{figure}

\begin{figure}[h]
\caption{Differential $D \to K^*$ distribution from
Refs.~\protect\cite{esixninekstar} and \protect\cite{esixfivekstar}.
The line is a fit used to extrapolate to $w=1$. The method of
extracting this plot from the published data is described in the text.}
\label{fig:getalpha}
\end{figure}

\begin{figure}[h]
\caption{Comparison of the leading order heavy quark predictions with
	data for the four measured $D$ meson semileptonic decay form factors.
	The solid and dotted lines correspond to the central value of
	$\alpha_s(\mu)$ obtained in
	\protect\subsecref{sec:EPD}{subsec:EPD-getalpha} and the one
	standard deviation uncertainties, respectively. The shaded bands
	correspond to the experimental values with one standard deviation
	uncertainties.}
\label{fig:ff}
\end{figure}

\begin{figure}[h]
\caption{Values of form factors associated with subleading operators
	obtained from experiment. The light and dark regions correspond to
	taking $\lambar = 200 \mbox{ and }400$ MeV, respectively. The light
	regions have been extended horizontally for clarity in some cases.}
\label{fig:Psi}
\end{figure}

\end{document}